# Evaluation of Boltzmann's *H*-function for Particles with Orientational Degrees of Freedom

**Shubham Kumar and Biman Bagchi**[*]

**Solid State and Structural Chemistry Unit, Indian Institute of Science, Bengaluru-560012, Karnataka, India**

[*]Email: profbiman@gmail.com; bbagchi@iisc.ac.in

## Abstract

*Boltzmann's H-function H(t), often regarded as an analog of time-dependent entropy, holds a venerable place in the history of science. However, accurate numerical evaluation of H(t) for particles other than atoms is rare. To remove this lacuna, we generalize Boltzmann's H-function to a gas of molecules with orientational degrees of freedom and evaluate H(t) from time-dependent joint probability distribution function f* (**p**, **L**, *t*) *for linear* (**p**) *and angular* (**L**) *momenta, evolving from an initial nonequilibrium state, by molecular dynamics simulations. We consider both prolate and oblate-shaped particles, interacting via well-known Gay-Berne potential and obtain the relaxation of the generalized molecular H(t) from initial (t = 0) nonequilibrium states. In the long-time limit, the H function saturates to its exact equilibrium value, which is the sum of translational and rotational contributions to the respective entropies. Both the translational and rotational components of H(t) decay nearly exponentially with time; the rotational component is more sensitive to the molecular shape that enters through the aspect ratio. A remarkable rapid decrease in the rotational relaxation time is observed as the spherical limit is approached, in a way tantalizingly reminiscent of Hu-Zwanzig hydrodynamic prediction with slip boundary condition. Additionally, we obtain H(t) analytically by solving the appropriate translational and rotational Fokker-Planck equation and obtain a modest agreement with simulations. We observe a remarkable signature of translation-rotation coupling as a function of molecular shape, captured through a physically meaningful differential term that quantifies the magnitude of translation-rotation coupling.*

# I. INTRODUCTION

In 1872, Ludwig Boltzmann introduced his *H*-function and *H*-theorem. The latter is one of the most celebrated theorems of physical science. It paved the way for the development of nonequilibrium statistical mechanics.[1–10] Boltzmann was the first to attempt to establish a relation between relaxation and entropy.[11] Even after 150 years, Boltzmann's *H*-function continues to play an important role in the study of nonequilibrium relaxation phenomena, partly because of its close relationship with entropy. Boltzmann's *H*-function continues to find wide use today, often with suitable generalization, as introduced by Kubo and employed by many.[7,12,13]

Boltzmann's *H*-function $H(t)$, in its original form, is defined by the following *D*-dimensional integration[1–10]:

$$H(t) = -\int d\mathbf{p}\, f(\mathbf{p},t) \ln f(\mathbf{p},t). \quad (1)$$

Here, **p** is the linear momentum of the particle, defined as $\mathbf{p} = m\mathbf{v}$; *m* is the mass of the particle and **v** is the velocity of the particle. $f(\mathbf{p},t)$ is the time-dependent linear momentum distribution function. Note that Boltzmann considered only translational degrees of freedom in three dimensions and considered only spherical atoms. The constraint to spheres has remained a limitation because the rotational degree of freedom could play an important, even dominant, role in many relaxation processes. Boltzmann's *H*-function (as defined by Eq. (1)) is the same as the Shannon entropy introduced in 1948 and is widely used in information theory, where a maximum entropy principle is often employed.[14–16] According to this principle, the distribution that maximizes entropy is selected as the true distribution.[17,18] In its popular form, Shannon entropy involves a sum over the probabilities of the observed states.

Boltzmann's *H*-theorem states that $H(t)$ is an ever-increasing function of time till it reaches the equilibrium value, which is essentially the same as the entropy for a dilute gas.[1–10] The *H*-theorem is stated as

$$\frac{dH}{dt} \geq 0. \qquad (2)$$

Here, the equality sign is satisfied only at equilibrium when the distribution takes on the Maxwell form. That is, if $f(\mathbf{p},t)$ is not an equilibrium distribution, the *H*-function will increase until the equilibrium distribution is obtained.

$H(t)$, in various forms, is widely regarded as a function to describe the time dependence of a nonequilibrium relaxation process. However, most of the discussions of Boltzmann's *H*-theorem and *H*-function center around the Boltzmann kinetic equation (also known as the Boltzmann transport equation), which is strictly applicable only to a dilute gas.[3–10] Originally proposed for monatomic gases, the Boltzmann kinetic equation was later extended to encompass rough spheres, loaded spheres, spherocylinders, polyatomic gases, real gases, and granular materials.[6,19–37] Furthermore, it was expanded to accommodate the quantum treatment of molecules with internal degrees of freedom, resulting in the Waldmann-Snider equation,[38,39] allowing for consideration of possible excitation of quantum superposition states and providing explanations for magnetic and electric field effects on gas transport coefficients.[21] It is well-known that Boltzmann's kinetic equation has been used in various areas, from astrophysics to hydrodynamics to chemical kinetics, and remains a venerable subject of classroom studies and academic discussions, not just in research. However, despite its importance, we find limited, precise numerical (or analytical) calculations of the evolution of Boltzmann's *H*-function from a well-defined nonequilibrium state.[8,12,13,40–43] This has long remained a lacuna in the area of time-dependent statistical mechanics. In fact, we are not aware of any analytical form of $H(t)$. The latter lacuna has recently been removed in Ref. 44.

Recently, we carried out a detailed numerical evaluation of $H(t)$ for a system of particles interacting via (i) hard sphere, (ii) soft sphere, and (iii) Lenard-Jones potentials.[44] We created different nonequilibrium states and studied the time evolution of the $H$-function in one, two, and three dimensions by calculating $f(\mathbf{p},t)$ itself. It was observed that Boltzmann's $H$-function captured many of the nuances of the relaxation phenomena. For example, in the one-dimensional Lennard-Jones system, it captured a correlated motion with a long lifespan, originating from single file motion. Another interesting result was a clear demonstration of the linear response of the system to initial perturbation, even when the distribution function at intermediate times appeared quite different.

All the reported theoretical and computer simulation studies of $H(t)$ employed spherical atoms/molecules with radially symmetric potentials,[8,12,13,40–44] despite the prevalent non-sphericity of real molecular systems. Many experimental studies often pertain to rotational dynamics, thus emphasizing the significance of the departure from spherical symmetry in molecular systems.[45] Furthermore, aspherical molecules offer a higher packing density compared to spherical ones,[46,47] emphasizing the increasing importance of molecular shape and geometry in determining physical properties.[48–51] This difference is particularly noteworthy as non-spherical molecules can rotate to facilitate molecular motion.[51,52] Such translational-rotational coupling is absent in spherical molecules, emphasizing the need for further attention to this additional channel.

In this work, we present the generalized form of Boltzmann's $H$-function for particles with orientational degrees of freedom (ODOF). Furthermore, we delve into the time evolution of the generalized $H$-function for prolate and oblate-shaped particles. The simulation results are compared with analytical forms that we derive by solving the Fokker-Planck (FP) equation of motion and using the resulting expression to obtain the $H(t)$.

The organization of the rest of the paper is as follows. In Section II, we present the generalized form of Boltzmann's *H*-function (Eq. 5). Section III contains the simulation and the system details. Here, we describe the construction of the initial nonequilibrium state. In Section IV, we present the numerical results. First of all, we discuss the time evolution of the generalized *H*-function for prolate and oblate-shaped particles. Subsequently, we discuss the aspect ratio dependence of the relaxation time of *H*(*t*). After that, we introduce a physically motivated differential term that captures the presence and the magnitude of translation-rotation coupling through a discriminant $\Delta_{T-R}(t)$. This function shows interesting dynamical behavior at intermediate times. Section V presents our analytical expression of *H*(*t*). In Section VI, we present a concise discussion of the results along with concluding remarks in connection with future problems.

## II. GENERALIZATION OF BOLTZMANN'S *H*-FUNCTION TO INCORPORATE ORIENTATION

For particles with orientational degrees of freedom (ODOF), Boltzmann's kinetic equation needs to be modified.[6,19–37] We now need to deal with the time-dependent joint probability distribution $f(\mathbf{r},\mathbf{p},\mathbf{\Omega},\mathbf{L},t)$ for finding an *aspherical particle* at position **r** with linear momentum **p**, having an orientation $\mathbf{\Omega}$ and angular momentum **L** (defined as $\mathbf{L}=I\mathbf{\omega}$; $\mathbf{\omega}$ is the angular velocity of the particle with the moment of inertia $I$). We then write Boltzmann's kinetic equation generalized to include orientation as

$$\left(\frac{\partial f}{\partial t}\right)_{coll} = \iiint d\mathbf{p}_2 d\mathbf{\Omega}_2 d\mathbf{L}_2 \, d\mathbf{p}'_1 d\mathbf{\Omega}'_1 d\mathbf{L}'_1 \, d\mathbf{p}'_2 d\mathbf{\Omega}'_2 d\mathbf{L}'_2 \, \delta(\mathbf{P}-\mathbf{P}') \\ \times \delta(\mathbf{L}_T - \mathbf{L}'_T)\delta(E-E')|T_{fi}|^2 \left(f^{(2)}_{1'2'} - f^{(2)}_{12}\right), \quad (3)$$

where the collisional integral is given in terms of two-particle joint probability distribution. Here, **P**, $\mathbf{L}_T$, and $E$ denote the total linear momentum, total angular momentum, and total energy before a collision, respectively; $\mathbf{P}'$, $\mathbf{L}'_T$, and $E'$ represent the total linear momentum, total angular momentum, and the total energy after the collision. The *T*-matrix gives the transition probability. $f^{(2)}$ is the two-body joint probability distribution function. $\mathbf{p}_1$ and $\mathbf{p}_2$ represent the linear momenta, $\mathbf{L}_1$ and $\mathbf{L}_2$ denote the angular momenta, and $\mathbf{\Omega}_1$ and $\mathbf{\Omega}_2$ represent the orientation of any two particles before a collision. After the collision, their linear momenta are represented by $\mathbf{p}'_1$ and $\mathbf{p}'_2$, their angular momenta by $\mathbf{L}'_1$ and $\mathbf{L}'_2$, and their orientations by $\mathbf{\Omega}'_1$ and $\mathbf{\Omega}'_2$. The expression contains the conservation of energy, linear momenta, and angular momenta. In Eq. (3), Boltzmann's collision integral is expressed in its modern form by introducing a transition matrix that we often employ in describing the form of the structure factor measured by neutron scattering experiments. However, it is the same as the original expression.

As experienced by Evans and coworkers,[53,54] the inclusion of the rotational degrees of freedom results in an even more complex equation than the original Boltzmann kinetic equation. It is highly non-trivial. In order to proceed further towards the appropriate *H*-function, we need to recognize the following things. In deriving the binary collisional event, one assumes that the collision occurs at a small volume element at position $\mathbf{r}$, where only the exchange of momenta occurs without any change of position. We proceed along the same line and assume that only the angular momenta are exchanged during collision, and no change in orientation occurs. *It is interesting to note that this is similar to the set of simplifying assumptions made by Gordon in the study of magnetic relaxation and is well-known as the J-diffusion model.*[55] However, here, one assumes that the momenta and the angular momenta change on the faster time scales. The collisional integral then reduces to a simpler form as

$$\left(\frac{\partial f}{\partial t}\right)_{coll} = \iiint d\mathbf{p}_2 d\mathbf{L}_2 \, d\mathbf{p}'_1 d\mathbf{L}'_1 \, d\mathbf{p}'_2 d\mathbf{L}'_2 \, \delta(\mathbf{P}-\mathbf{P}')\delta(\mathbf{L}_T - \mathbf{L}'_T) \\ \times \delta(E - E')|T_{fi}|^2 \left(f^{(2)}_{1'2'} - f^{(2)}_{12}\right). \quad (4)$$

This expression is tractable in the same way as Boltzmann's original kinetic equation without orientation. In the next step, one employs the assumption of molecular chaos to decompose the two-particle joint probability distribution $f^{(2)}(\mathbf{p}_1,\mathbf{p}_2,\mathbf{L}_1,\mathbf{L}_2,t)$ into the product of two single-particle distribution functions $f(\mathbf{p},\mathbf{L},t)$. This is the assumption of molecular chaos. In the subsequent steps, one needs to assure that such an approximate distribution function indeed attains the Maxwell velocity distribution in a long time. The steps are the same as documented in textbooks.[8,10] This necessitates the introduction of Boltzmann's *H*-function, now generalized to include the contribution of angular momentum (**L**). The generalized or total Boltzmann's *H*-function $H_{Tot}(t)$ is given by

$$H_{Tot}(t) = -\int d\mathbf{p}\, d\mathbf{L}\, f(\mathbf{p},\mathbf{L},t) \ln f(\mathbf{p},\mathbf{L},t). \qquad (5)$$

We note that the joint probability distribution *may not* be decomposed into a product of independent translational and rotational distribution functions. This is an important issue to be discussed in detail further below.

However, even for non-spherical particles, it is still useful to define translational and rotational *H*-functions. The translational part has already been defined by Eq. (1). The rotational part is defined in the same fashion as

$$H_{Rot}(t) = -\int d\mathbf{L}\, f(\mathbf{L},t) \ln f(\mathbf{L},t), \qquad (6)$$

where $f(\mathbf{L},t)$ is the time-dependent angular momentum distribution function.

The intermolecular potential inherently involves translation-rotation coupling, making it impossible to decompose the joint probability distribution into the product of two independent functions. This is an important point. Because of the exchange of momentum and kinetic energy between the translational and rotational degrees of freedom, such a decomposition can lead to serious error, as indeed demonstrated in this work.

This result, in fact, reminds one of the assumptions of molecular chaos all over again, but in a different context, at the level of the single-particle joint probability distribution, while the original molecular chaos assumption involves the decomposition of two-particle distribution into the product of two one-particle distribution functions. In a long time, that is, the equilibrium limit, the translational, and the rotational momenta are decoupled, and the function gives the total kinetic entropy of the gas. During the evolution of an initial nonequilibrium distribution function, $f(\mathbf{r},\mathbf{p},\mathbf{\Omega},\mathbf{L},t=0)$, it is non-trivial to obtain the time-dependent joint probability distribution $f(\mathbf{r},\mathbf{p},\mathbf{\Omega},\mathbf{L},t)$.

In fact, theoretically, one can construct a generalized Boltzmann's kinetic equation with arbitrary internal degrees of freedom and follow the same steps as Boltzmann to arrive at Eq. (5) for the *H*-function. Such a definition must follow all the necessary conditions (like conservation laws and transition probabilities between states) and can serve the same purpose. Indeed, Kubo *et al.* introduced an *H*-function involving a generalized coordinate-dependent density function given by[7]

$$H_{Kubo}(t) = -\int dx P(x,t) \ln\left(\frac{P(x,t)}{P_{eq}(x)}\right) \qquad (7)$$

This generalized coordinate-dependent form of *H*-function serves a similar purpose as the original. Here, $P_{eq}(x)$ denotes the equilibrium distribution of a given variable *x*. Typically, *x* is assumed to represent a position variable or a combination thereof. This formulation aligns with the approach discussed in influential works, such as the monograph by Kubo and Toda,[7] though it does not encompass any orientation-dependent functions.

To comprehend the time evolution of the generalized *H*-function, as defined by Eq. (5), and to check the validity of the *H*-theorem, we investigate two model systems featuring rotational degrees of freedom. The time evolution of the generalized *H*-function is evaluated for symmetrical ellipsoids (ellipsoids of revolution). To be more specific, we consider a system of gas consisting of prolates and oblates and vary the aspect ratio from markedly non-spherical to spherical limits. The simulation results are compared with analytical forms that we derive by solving the Fokker-Planck equation of motion and using the resulting expression to obtain the $H(t)$.

Our studies reveal several interesting new results. We find that the normalized rotational and translational $H(t)$ grows with similar time scales from the initial nonequilibrium state. *However, their aspect ratio dependencies are strikingly different*. In the following section, we describe the simulation details.

## III. SIMULATION DETAILS

We have performed extensive nonequilibrium molecular dynamic simulations of dilute gases with orientational degrees of freedom to study the evolution of the generalized $H$-function, defined by Eq. (5). Our model system consists of 4000 symmetrically ellipsoidal particles, i.e., prolates (rod-shaped molecules) or oblates (disc-shaped molecules) contained in a cubic box having the usual periodic boundary condition in each case. We have carried out these simulations in the microcanonical ensemble (constant N, V, and E). In simulation studies, where the shape and orientation of the rigid body play a crucial role, the Gay-Berne (GB) pair potential provides a convenient model for interaction between two particles.[51,56–59] In the GB pair potential, each spheroid $i$ is represented by the position $\mathbf{r}_i$ of its center of mass and a unit vector $\mathbf{e}_i$ along the principal symmetry axis (as shown in **Figure 1**). The Gay-Berne potential for the interaction between two spheroids $i$ and $j$ is given by [56]

$$U_{ij}^{GB}\left(\mathbf{r}_{ij}, \mathbf{e}_i, \mathbf{e}_j\right) = 4\varepsilon_{ij}\left(\hat{\mathbf{r}}_{ij}, \mathbf{e}_i, \mathbf{e}_j\right)\left(\rho_{ij}^{-12} - \rho_{ij}^{-6}\right), \qquad (8)$$

where

$$\rho_{ij} = \frac{r_{ij} - \sigma(\hat{\mathbf{r}}_{ij}, \mathbf{e}_i, \mathbf{e}_j) + \sigma_0}{\sigma_0}. \tag{9}$$

Here, $\sigma_0$ defines the cross-sectional diameter along the breadth, $r_{ij}$ is the distance between the two centers of mass, and $\hat{\mathbf{r}}_{ij} = \mathbf{r}_{ij}/r_{ij}$ is a unit vector along the intermolecular separation vector $\mathbf{r}_{ij}$. The molecular shape parameter $\sigma(\hat{\mathbf{r}}_{ij}, \mathbf{e}_i, \mathbf{e}_j)$ is given by

$$\sigma(\hat{\mathbf{r}}_{ij}, \mathbf{e}_i, \mathbf{e}_j) = \sigma_0 \left[ 1 - \frac{\chi}{2} \left\{ \frac{(\mathbf{e}_i \cdot \hat{\mathbf{r}}_{ij} + \mathbf{e}_j \cdot \hat{\mathbf{r}}_{ij})^2}{1 + \chi(\mathbf{e}_i \cdot \mathbf{e}_j)} - \frac{(\mathbf{e}_i \cdot \hat{\mathbf{r}}_{ij} - \mathbf{e}_j \cdot \hat{\mathbf{r}}_{ij})^2}{1 - \chi(\mathbf{e}_i \cdot \mathbf{e}_j)} \right\} \right]^{-1/2}, \tag{10}$$

where $\chi = \dfrac{(\kappa^2 - 1)}{(\kappa^2 + 1)}$. Here $\kappa$ denotes the aspect ratio of the spheroid and is given by $\kappa = \sigma_E/\sigma_S$. $\sigma_E$ is the molecular length along the principal symmetry axis and $\sigma_S = \sigma_0$.

The energy parameter $\varepsilon(\hat{\mathbf{r}}_{ij}, \mathbf{e}_i, \mathbf{e}_j)$ is given by

$$\varepsilon_{ij}(\hat{\mathbf{r}}_{ij}, \mathbf{e}_i, \mathbf{e}_j) = \varepsilon_0 \left[ \varepsilon_1(\mathbf{e}_i, \mathbf{e}_j) \right]^\nu \left[ \varepsilon_2(\hat{\mathbf{r}}_{ij}, \mathbf{e}_i, \mathbf{e}_j) \right]^\mu, \tag{11}$$

where $\mu$ and $\nu$ are two exponents that are adjustable and

$$\varepsilon_1(\mathbf{e}_i, \mathbf{e}_j) = \left[ 1 - \chi^2(\mathbf{e}_i \cdot \mathbf{e}_j)^2 \right]^{-1/2}, \tag{12}$$

and

$$\varepsilon_2(\hat{\mathbf{r}}_{ij}, \mathbf{e}_i, \mathbf{e}_j) = 1 - \frac{\chi'}{2} \left[ \frac{(\mathbf{e}_i \cdot \hat{\mathbf{r}}_{ij} + \mathbf{e}_j \cdot \hat{\mathbf{r}}_{ij})^2}{1 + \chi'(\mathbf{e}_i \cdot \mathbf{e}_j)} + \frac{(\mathbf{e}_i \cdot \hat{\mathbf{r}}_{ij} - \mathbf{e}_j \cdot \hat{\mathbf{r}}_{ij})^2}{1 - \chi'(\mathbf{e}_i \cdot \mathbf{e}_j)} \right]. \tag{13}$$

Here, $\chi' = (\kappa'^{1/\mu} - 1)/(\kappa'^{1/\mu} + 1)$ with $\kappa' = \varepsilon_S/\varepsilon_E$. $\varepsilon_S$ represents the depth of the minimum of the potential for a pair of ellipsoids when they are aligned side-by-side $[\varepsilon_S = \varepsilon_0]$, and $\varepsilon_E$ is the corresponding depth for end-to-end alignment. The functional form of the Gay-Berne potential and a schematic diagram of a pair of ellipsoids interacting via the Gay-Berne potential is shown in **Figure 1**.

All quantities are given here in reduced units, defined in terms of the GB potential parameters $\sigma_0$ and $\varepsilon_0$: length in the units of $\sigma_0$, time in the units of $(\sigma_0^2 m/\varepsilon_0)^{1/2}$, $m$ being the mass of the spheroid, and temperature in units of $\varepsilon_0/k_B$, $k_B$ being the Boltzmann constant. $\varepsilon_0$ has been taken to unity. We have set the mass ($m$) of the ellipsoids equal to unity, i.e., $m = 1.0$. The inertia tensor is chosen as $I_x = I_y = 1.0$. The value of the $I_z$ component is irrelevant due to the conservation of the angular velocity along the symmetry ($z$) axis of the ellipsoid.[60] The reduced density $\left(\rho^* = \frac{N \times m}{V} \times \frac{4\pi}{3m} \times \frac{\sigma_S^2 \sigma_E}{8}\right)$ is taken as 0.10 for all the systems, where $N$ is the total number of particles, and $V$ is the volume of the simulation box.

The Gay-Berne potential encompasses a range of potentials, each distinguished by a unique set of four parameters $(\kappa, \kappa', \mu \text{ and } \nu)$. The parameter $\kappa$, indicating the aspect ratio, quantifies shape anisotropy, while $\kappa'$ measures the anisotropy of the well depth. This latter aspect can be adjusted using the parameters $\mu$ and $\nu$. In our investigation, we characterize the spheroid using these parameters as $(\kappa, 5, 2 \text{ and } 1)$. The aspect ratio $(\kappa)$ for prolates is varied from 1.05 to 3.0, whereas, for oblates, it is varied from 0.98 to 0.40.

The initial configurations (position) have been taken from equilibrium simulations in the canonical (NVT) ensemble corresponding to $T^* = 2.0$. Following this, the initial nonequilibrium state in linear momentum space is created by setting the amplitude of linear velocities for all particles equally. This magnitude aligns with the equipartition theorem corresponding to the reduced temperature $T^* = k_B T/\varepsilon_0 = 2.0$. A similar approach is utilized to generate the initial nonequilibrium state in angular momentum space. This approach assures that the total energy remains constant.

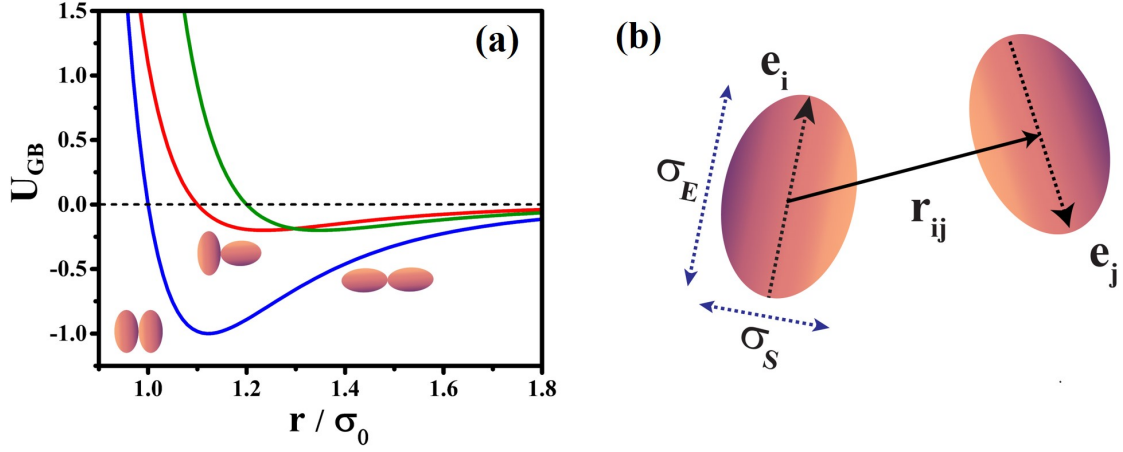

Figure 1: (a) Functional form of the Gay-Berne potential characterized by a set of parameters ($\kappa$, 5, 2, 1). The ratio of the energy depth of the side-by-side configuration (corresponding to the deepest energy depth, blue line) to that of the end-to-end configuration (corresponding to the shallowest energy depth, green line) shows the value of $\kappa' = 5.0$. (b) Schematic diagram of a pair of ellipsoids defined by the Gay-Berne potential parameter.

The equations of motion have been integrated following the velocity-Verlet algorithm[61,62] with the integration time step of $\delta t^* = 0.001\,\tau$. Here $\tau$ represents the scaled time, defined as $\tau = \sigma_0 \sqrt{m/\varepsilon_0}$. In this work, we have converted the time scales from reduced time ($t^*$) to real-time (t) using relation $t^* = t\sqrt{\dfrac{\varepsilon_0}{m\sigma_0^2}}$, where the values of $\varepsilon_0$, $m$ and $\sigma_0$ have been taken corresponding to that of argon atom, i.e., $\varepsilon_0/k_B = 119.8\,K$, $m = 0.03994\ kg/mol$ and $\sigma_0 = 3.405\times 10^{-10}\,m$.

From molecular dynamics simulations, starting from well-defined nonequilibrium states, we have meticulously tracked the temporal evolution of the components of linear momentum ($p_x$, $p_y$, $p_z$) and the components of angular momentum ($L_x$, $L_y$) for each particle. To compute the translational H-function, $H_{trans}(t)$, we have constructed three-dimensional grids corresponding to $p_x$, $p_y$, and $p_z$. These grids capture the evolution of the probability distribution function $f(\mathbf{p},t)$, where $\mathbf{p} = (p_x, p_y, p_z)$, over time. Subsequently, we have performed numerical integration over these grids to evaluate the translational $H$-function, $H_{trans}(t) = -\iiint dp_x\, dp_y\, dp_z\, f(\mathbf{p},t) \ln f(\mathbf{p},t)$. For the rotational $H$-function, $H_{Rot}(t)$, we have utilized two-dimensional grids corresponding to $L_x$ and $L_y$, capturing the time evolution of the probability distribution function $f(\mathbf{L},t)$, where $\mathbf{L} = (L_x, L_y)$. The rotational $H$-function is computed by the numerical integration over these grids as: $H_{Rot}(t) = -\iint dL_x dL_y\, f(\mathbf{L},t) \ln f(\mathbf{L},t)$. The generalized $H$-function, $H_{Tot}(t)$, considers the combined dynamics of translational and rotational degrees of freedom. Therefore, we extend our approach to five-dimensional grids encompassing $p_x$, $p_y$, $p_z$, $L_x$ and $L_y$, allowing us to comprehensively capture the time evolution of the probability distribution function $f(\mathbf{p},\mathbf{L},t)$. Subsequently, we have performed numerical integration to obtain $H_{Tot}(t)$. We have conducted five independent simulations in order to achieve the converged results, and the outcomes are averaged over all of these runs.

In the following sections, we present the numerical results obtained from our simulations and compare them with theoretical closed-form expressions derived from the Fokker-Planck equations.

## IV. RESULTS AND DISCUSSION

We study the time evolution of the generalized (or total) $H$-function, defined by Eq. (5), using a series of nonequilibrium molecular dynamics simulations of dilute gases with orientational degrees of freedom (ODOF) interacting via the Gay-Berne pair potential. In the following subsections, we present the numerical results obtained from our simulations and compare them with theoretical closed-form expressions derived from the Fokker-Planck (FP) equations.

### (A) Time dependence of *H*-function

In **Figures 2 (a) and 2 (b)**, we show the time evolution of the generalized (or total) Boltzmann *H*-function $H_{Tot}(t)$, as well as its translational and rotational components for oblate $(\kappa = 0.8)$ and prolate $(\kappa = 2.0)$, respectively, at reduced density, $\rho^* = 0.10$ and the average reduced temperature, $T^*=2.0$. We observe that for both oblates and prolates, the generalized Boltzmann *H*-function, as well as its translational and rotational components, sharply increase in a short time, followed by a subsequently monotonous approach to the equilibrium value (shown by dotted lines) at a longer time, which is the equilibrium value at the chosen temperature. Further, we find that, in the long-time limit, the generalized Boltzmann *H*-function $H_{Tot}(t)$ saturates to the equilibrium value, which is the sum of translational and rotational contributions. However, at intermediate times, the sum of translational and rotational contributions is not equivalent to the generalized Boltzmann *H*-function $H_{Tot}(t)$. It can be attributed to the translational-rotational coupling in the ellipsoids. That is, the time-dependent joint probability distribution is not decomposable into the product of two terms, as is possible under equilibrium conditions.

To get the relaxation times of the generalized Boltzmann *H*-function and its translational and rotational parts, the respective *H*-functions are fitted with the functional form $H(t) = H_{eq} + \left[\left(H(0) - H_{eq}\right)\exp(-t/\tau)\right]$. In **Figures 2 (c)** and **2 (d),** we show normalized *H(t)*, defined as $\left(\left(H(t) - H_{eq}\right)/\left(H(0) - H_{eq}\right)\right)$ and its exponential fit for oblate $(\kappa = 0.8)$ and prolate $(\kappa = 2.0)$, respectively. The relaxation times $(\tau)$ for the generalized, translational, and rotational H-functions are given in **Table 1**. We observe that the relaxation time of the generalized *H*-function lies in between the relaxation times of the translational and rotational components.

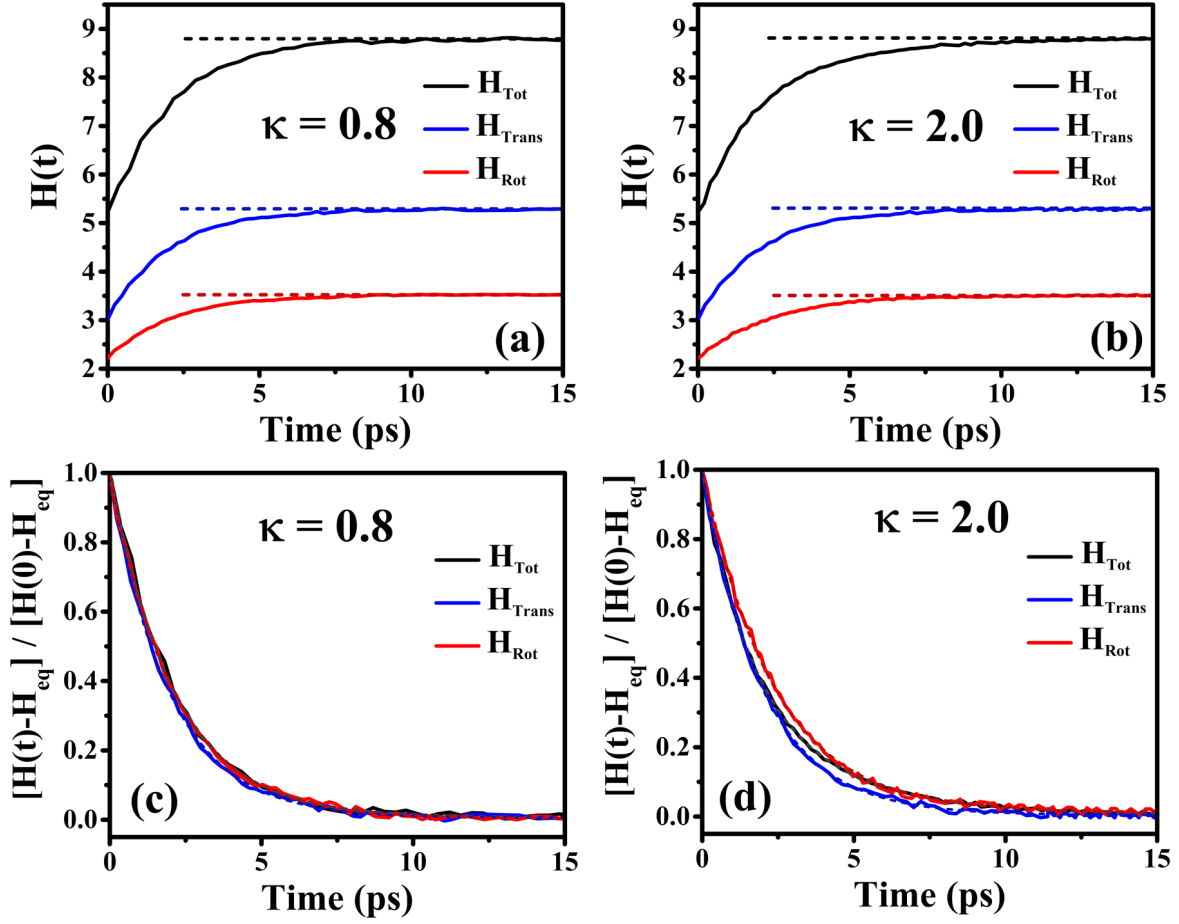

**Figure 2:** Time evolution of the generalized, translational, and rotational *H*-functions obtained via computer simulations for (a) oblate ($\kappa = 0.8$) and (b) prolate ($\kappa = 2.0$) systems of dilute gases (at reduced density $\rho^* = 0.10$, and average reduced temperature, $T^*=2.0$) interacting with Gay-Berne potential. In all the cases, the *H*-function increases monotonically and then attains equilibrium at a longer time, which is the equilibrium value at the final temperature, shown by corresponding dashed lines in the figures. Panels (c) and (d) depict the time evolution of the normalized *H*(*t*) [for the results shown in panels (a) and (b), respectively]. The corresponding dotted lines show the exponential fitting of the normalized *H*(*t*).

It is to be noted that in the present case, the rotational contribution to the final equilibrium value of $H_{Tot}$ is approximately half that of the translational contribution across all $\kappa$ values examined. This observation arises from the inherent balance between translational and rotational degrees of freedom, as well as the mass and moment of inertia of particles, which fundamentally influence the relative contributions of translational and rotational dynamics, as later shown by Eqs. (17) and (18). Additionally, for water molecules in the liquid phase, the rotational contribution is observed to be significant but smaller by a similar proportion than the translational component.[63,64] In view of the present observation, it would be interesting to calculate the *H*-function for real molecules like water, carbon dioxide, and so forth.

**Table 1: The relaxation times obtained by the exponential fitting of the generalized, translational, and rotational *H*-functions for oblate ($\kappa = 0.8$) and prolate ($\kappa = 2.0$).**

| System | Relaxation Time (in ps) | | |
|---|---|---|---|
| | $\tau_{Tot}$ | $\tau_{Trans}$ | $\tau_{Rot}$ |
| **Oblate ($\kappa = 0.8$)** | 2.08 ± 0.03 | 1.96 ± 0.04 | 2.12 ± 0.01 |
| **Prolate ($\kappa = 2.0$)** | 2.14 ± 0.02 | 1.99 ± 0.03 | 2.34 ± 0.02 |

**(B) Aspect ratio dependence**

In order to get an insight into the effect of the aspect ratio of ellipsoids on the relaxation times of the *H*-function, we compute the *H*-functions for different aspect ratios of oblates and prolates. In **Figure 3 (a),** we show the variation in the relaxation time of the generalized Boltzmann *H*-function $H_{Tot}(t)$ as a function of the aspect ratio of ellipsoids. The variation of the relaxation times of translational and rotational *H*-functions with the aspect ratio of ellipsoids is shown in **Figure 3 (b).** It is evident from **Figure 3** that the nonequilibrium relaxation function *H*(t) is sensitive to the aspect ratio of the ellipsoids. The relaxation time gets slower as the particle becomes either more disc-like or rod-like. Further, we observe that rotational relaxation is more sensitive to the aspect ratio than translational relaxation time.

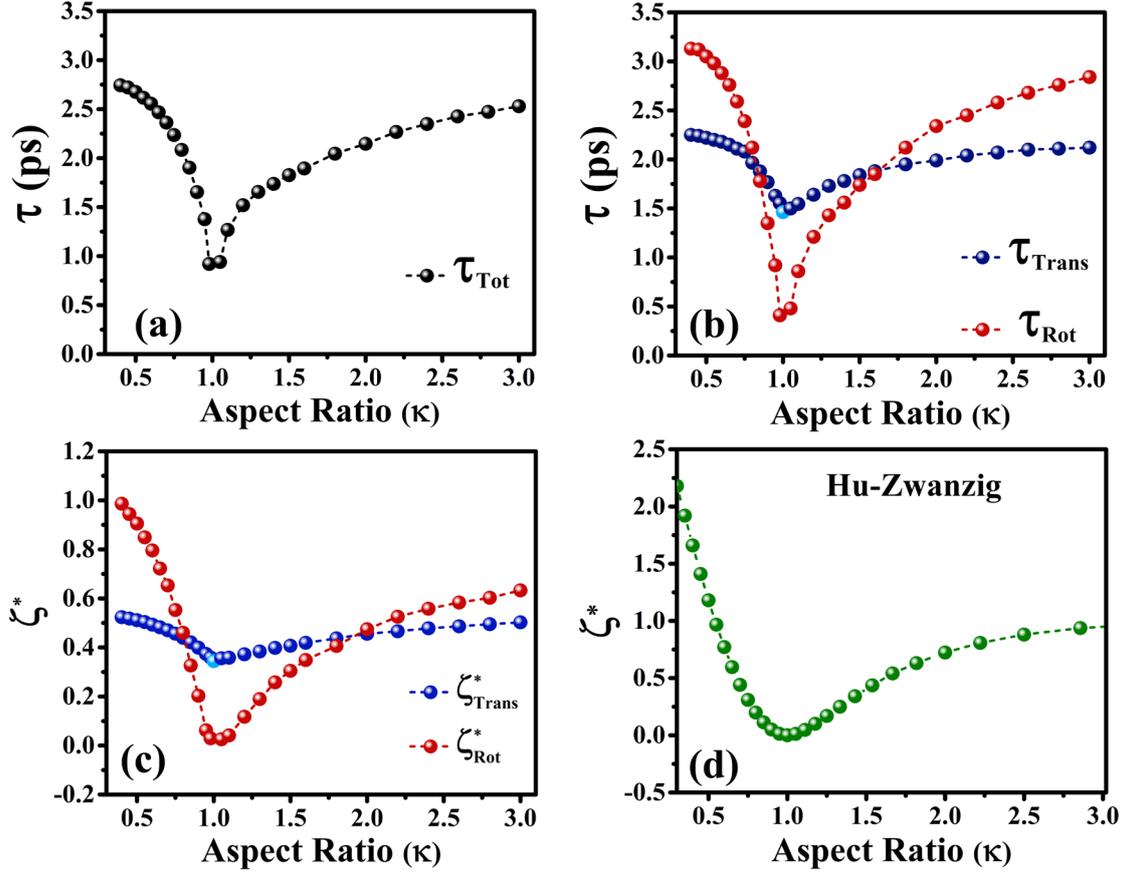

**Figure 3:** The variation of the relaxation time obtained by the exponential fit of (a) generalized and (b) translational and rotational *H*-functions with the aspect ratio of the ellipsoids. It is to be noted that the rotational relaxation is more sensitive to the aspect ratio. At $\kappa = 1$, the relaxation time of the translational *H*-function corresponds to the spherical particles interacting via Lennard-Jones potential (shown by the sky-blue symbol). (c) The variation of the translational $(\zeta_{Trans})$ and rotational $(\zeta_{Rot})$ frictions with the aspect ratio of the ellipsoids. As the spherical limit is approached, a rapid decrease in the rotational friction is observed, which is consistent with the hydrodynamic predictions of Hu and Zwanzig (panel (d)) with slip boundary condition. The data presented in panel (d) is sourced from Ref. [65].

The observed aspect ratio dependencies in the relaxation times of *H*-functions highlight the significant role of particle geometry and moment of inertia, where changes in aspect ratio influence rotational relaxation times more prominently than translational relaxation times. This sensitivity reflects the complex interplay between particle shape and internal dynamics. The differences in relaxation times between translational and rotational motions may be attributed to a 'bottleneck effect' in the phase space, as discussed by Shudo and coworkers[66,67] in the context of liquid water. This effect describes a phenomenon where one type of motion (either translational or rotational) experiences greater hindrance or restriction compared to the other due to constraints in phase space availability, influenced by moments of inertia and particle geometry.

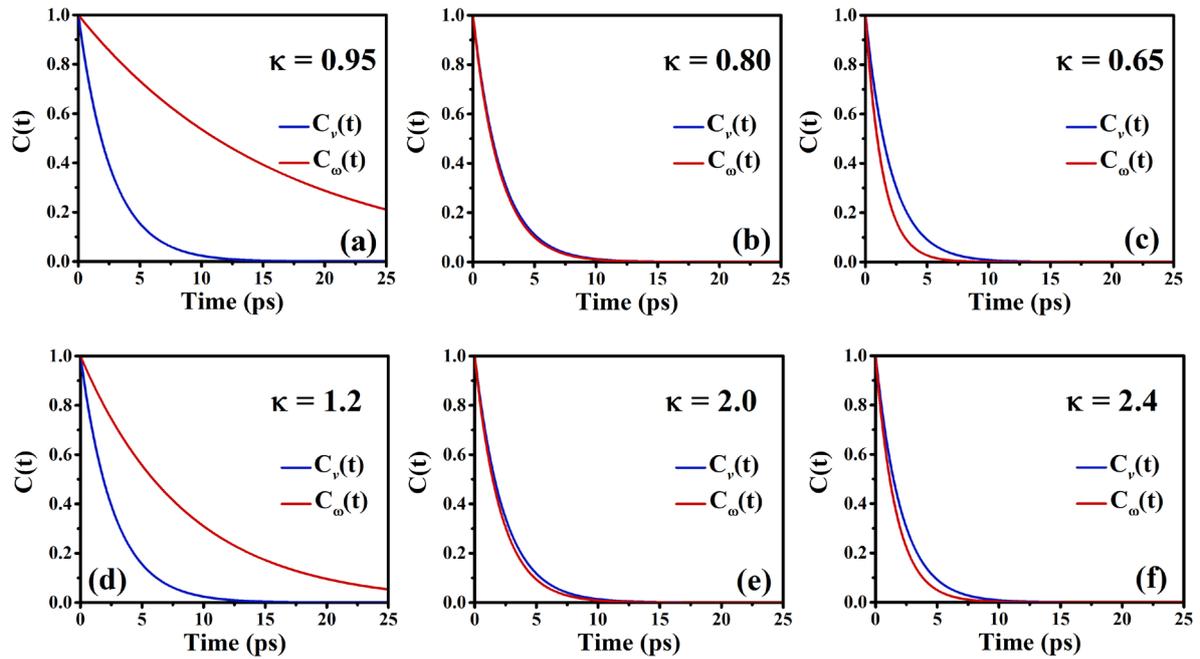

**Figure 4: The normalized linear and angular velocity autocorrelation functions of oblates when (a) $\kappa = 0.95$, (b) $\kappa = 0.80$, (c) $\kappa = 0.65$, and for prolates when (d) $\kappa = 1.2$, (e) $\kappa = 2.0$, and (f) $\kappa = 2.4$, respectively. It is to be noted that the decay of the angular velocity autocorrelation function becomes faster as the particle becomes either more disc-like or rod-like.**

To further understand the strong aspect ratio dependence of the relaxation times, we calculate the linear and angular velocity time-correlation functions as well as the corresponding frictions. **Figures 4 (a)** to **4 (c)** depict the normalized linear and angular velocity time-correlation functions of oblates with different aspect ratios $(\kappa = 0.95, 0.80 \text{ and } 0.65)$. The time evolution of the normalized linear and angular velocity time-correlation functions of prolates with three different aspect ratios $(\kappa = 1.2, 2.0 \text{ and } 2.4)$ are shown in **Figures 4 (d)** to **4 (f).** It is observed that when the aspect ratio of the ellipsoid is close to one, the decay of the angular velocity autocorrelation function is slower than that of the linear velocity autocorrelation function. However, the decay of the angular velocity autocorrelation function becomes faster as the particle becomes either more disc-like or rod-like. This aspect deserves further study.

We calculate translational friction $(\zeta_{Trans})$ via the integration of the linear velocity time-correlation function and using Einstein's relation $D_{Trans} = k_B T / \zeta_{Trans}$. Similarly, rotational friction $(\zeta_{Rot})$ is obtained via the angular velocity time-correlation function using Einstein's relation $D_{Rot} = k_B T / \zeta_{Rot}$.[45,68] In **Figure 3 (c)**, we show the variation of translational and rotational friction of ellipsoids as a function of the aspect ratio. We find that rotational friction $(\zeta_{Rot})$ shows strong aspect ratio dependence as observed for the relaxation times of rotational H-function. It is to be noted that the variation of the rotational friction with the aspect ratio behaves in a similar manner to that predicted by the hydrodynamic theory of Hu and Zwanzig (as shown in **Figure 3 (d)**), who obtained the friction for prolate and oblate ellipsoids under the slip boundary condition.[65] Such a dramatic behavior is not predicted for rotational friction under the stick boundary condition because the rotational friction itself is very large for sticky spheres.

**(C) Translation-rotation coupling**

In order to understand the aspect-ratio dependence of the translational and rotational relaxation times, we study translation-rotation coupling. To quantify the coupling between the translational and rotational motion of ellipsoids evolving from initial nonequilibrium condition, we define a quantity $\Delta_{T-R}(t)$ in the following manner:

$$\Delta_{T-R}(t) = \int d\mathbf{p}\, d\mathbf{L}\left[f(\mathbf{p},\mathbf{L},t) - f(\mathbf{p},t)f(\mathbf{L},t)\right]. \tag{14}$$

Here, the symbols have their usual meaning, as described earlier.

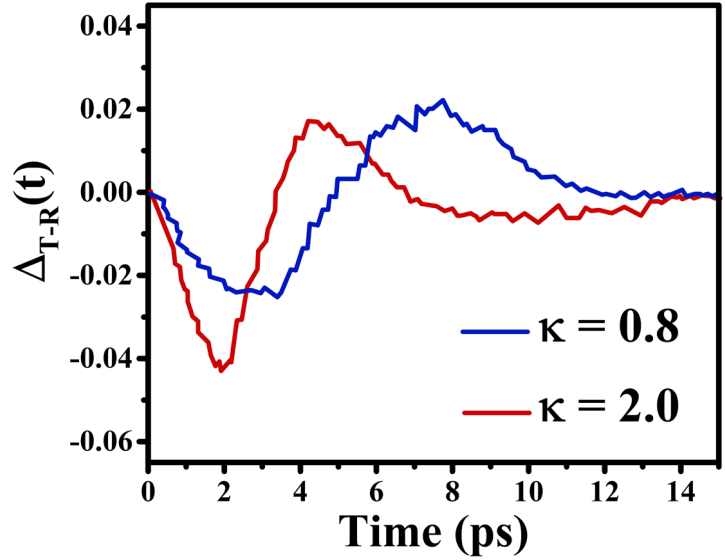

**Figure 5: A measure of translation-rotation coupling. Time evolution of $\Delta_{T\text{-}R}(t)$ for oblate ($\kappa$ = 0.8) and prolate ($\kappa$ = 2.0), when the system is evolving from a nonequilibrium state to the equilibrium state. At $t$ = 0, $\Delta_{T\text{-}R}(t)$ is zero as the initial state is chosen such that there is no coupling between the translational and rotational motion of the particles. At intermediate times, it acquires a non-zero value due to translational-rotational coupling and again becomes zero in the long-time limit due to decoupling.**

In **Figure 5,** we show the variation of the quantity $\Delta_{T-R}(t)$ as a function of time for oblate $(\kappa = 0.8)$ and prolate $(\kappa = 2.0)$ when the system is evolving from the initial nonequilibrium state to the equilibrium state. At $t$ = 0, the quantity $\Delta_{T-R}(t)$ is zero since the initial nonequilibrium state is chosen such that there is no coupling between the translational and rotational motion of the particles. However, at intermediate time steps, we find that $\Delta_{T-R}(t)$ acquires a non-zero value suggesting a significant coupling between the translational and rotational motion of the ellipsoids. Further, as the system approaches the equilibrium, $\Delta_{T-R}(t)$ becomes zero due to the decoupling between translational and rotational dynamics. It is to be noted that the function $\Delta_{T-R}(t)$ is dependent on the initial nonequilibrium conditions.

**(D) Moment of inertia dependence**

In the earlier sections, we have discussed the aspect ratio dependence of translational, rotational, and generalized $H(t)$, keeping the moment of inertia fixed. The next obvious question arises what would be the relaxation time of $H(t)$ if the moment of inertia of spheroids is varied, keeping the aspect ratio fixed. To examine this, we have obtained the time evolution of translational, rotational, and generalized H(t) corresponding to different moments of inertia of ellipsoids for two different fixed aspect ratios.

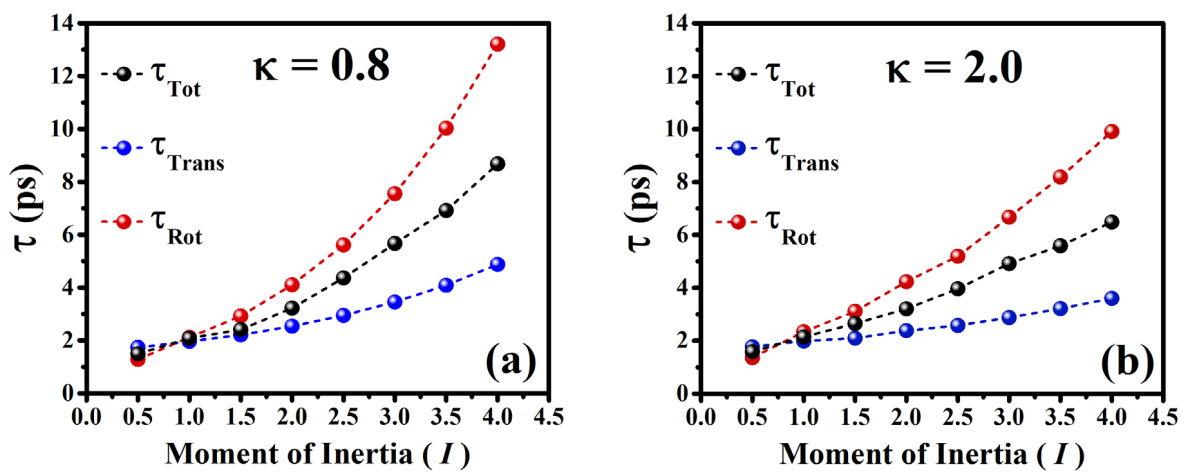

**Figure 6:** The variation of the relaxation time obtained by the exponential fit of generalized, translational, and rotational *H(t)* as a function of the moment of inertia of (a) oblate ($\kappa = 0.8$) and (b) prolate ($\kappa = 2.0$), for systems of dilute gases (at reduced density $\rho^* = 0.10$, and average reduced temperature, $T^*=2.0$). It is to be noted that as the moment of inertia increases, the relaxation of *H(t)* becomes slower.

In **Figures 6 (a)** and **6 (b)**, we show the variation in the relaxation time of the translational, rotational, and generalized H-functions as a function of the moment of inertial of oblate $(\kappa = 0.8)$ and prolate $(\kappa = 2.0)$, respectively. We observe that the relaxation time of the rotational *H*-function shows a strong dependence on the moment of inertia of ellipsoids for both prolates and oblates. As the moment of inertia increases, the relaxation of the rotational *H*-function becomes slower. Further, the relatively weak dependence of the translational relaxation time on the moment of inertia of ellipsoids can be attributed to translational-rotational coupling. As observed in earlier cases, the relaxation time of the generalized (or total) *H*-function lies in between that of the rotational and translation *H*-functions.

## V. ANALYTICAL EXPRESSION OF *H*(t)

A closed-form analytical expression for the H-function can be derived by using the solution of the Fokker-Planck (FP) equation of the single-particle momentum distribution.[44] The Fokker-Planck equation in the linear momentum space for $f(\mathbf{p},t)$ is given by[7,45]

$$\frac{\partial \delta f(\mathbf{p},t)}{\partial t} = \zeta_{Trans} \left( \frac{\partial}{\partial \mathbf{p}} \left[ \frac{\mathbf{p}}{m} + <E> \frac{\partial}{\partial \mathbf{p}} \right] \right) \delta f(\mathbf{p},t) \qquad (15)$$

where $<E>$ is the average energy and $\partial/\partial \mathbf{p}$ is the *D*-dimensional gradient in linear momentum space. Here $\delta f(\mathbf{p},t)$ is the deviation of the momentum distribution from the Maxwell distribution, which is the equilibrium state, i.e., $\delta f(\mathbf{p},t) = f(\mathbf{p},t) - f_M(\mathbf{p})$. In three-dimension, the above equation has the solution

$$f(\mathbf{p},t) = \frac{1}{(2\pi m k_B T (1-\Gamma^2(t)))^{3/2}} \exp(-[\mathbf{p} - \mathbf{p}_0 \Gamma(t)]^2 / (2m k_B T (1-\Gamma^2(t)))), \qquad (16)$$

where $\Gamma(t) = e^{-\zeta_{Trans} t}$, $\zeta_{Trans}$ being the translational friction of the system. Eq. (16) is solved with the initial condition that $f(\mathbf{p}, t=0) = \delta(\mathbf{p}-\mathbf{p}_0)\delta t$, where $\mathbf{p}_0$ is the initial momentum. It is essentially Green's function for the differential equation (15).

By putting Eq. (16) into Eq. (1), followed by some algebraic manipulations, one can obtain a closed-form analytic expression for translational H-function as[44]

$$H_{Trans}(t) = -\frac{3}{2}\ln\left(\frac{1}{2\pi m k_B T (1-\exp(-2\zeta_{Trans} t))}\right) + \frac{3}{2}. \qquad (17)$$

The above, rather elegant, expression of the translational H-function needs the input of translational friction $\zeta_{Trans}$.

Analogous to the closed-form analytical expression for the translational H-function, one can obtain an analytical expression for the rotational H-function using the Fokker-Planck equation for rotational motion. The closed-form analytical expression of rotational H-function for symmetrical ellipsoids is given by,

$$H_{Rot}(t) = -\ln\left(\frac{1}{2\pi I k_B T (1-\exp(-2\zeta_{Rot} t))}\right) + 1. \qquad (18)$$

The above expression of the rotational H-function needs the input of rotational friction $\zeta_{Rot}$.

The following remarks pertain to Eqs. (17) and (18). First, the respective frictions are obtained from fitting to the respective velocity time correlation functions. Both the linear and rotational velocity time correlation functions decay mostly exponentially, so extraction of the frictions by fitting to the solution of the ordinary Langevin equation is trivial. Alternatively, one can also obtain the frictions from respective diffusion coefficients by using the well-known Einstein relation $(D = k_B T/\zeta)$. Second, there is an issue with an unphysical ln(t) dependence at short times, or more precisely, in the limit as $t \to 0$. This unphysical behavior arises from our use of the Markovian approximation, which results in finite friction even at very short times, where friction should tend to zero. However, we have not yet been able to derive an expression of $H(t)$ that is valid under the non-Markovian limit.

In **Figure 7**, we compare the results obtained via simulation and the closed-form analytical expressions of translation and rotational $H$-function for oblate $(\kappa = 0.8)$ and prolate $(\kappa = 2.0)$. There is an interesting observation to be made here. For atomic systems, the expression of $H(t)$ from the Fokker-Planck equation provides almost a quantitative agreement with the simulated values, except at a very short time, as discussed above.[44]

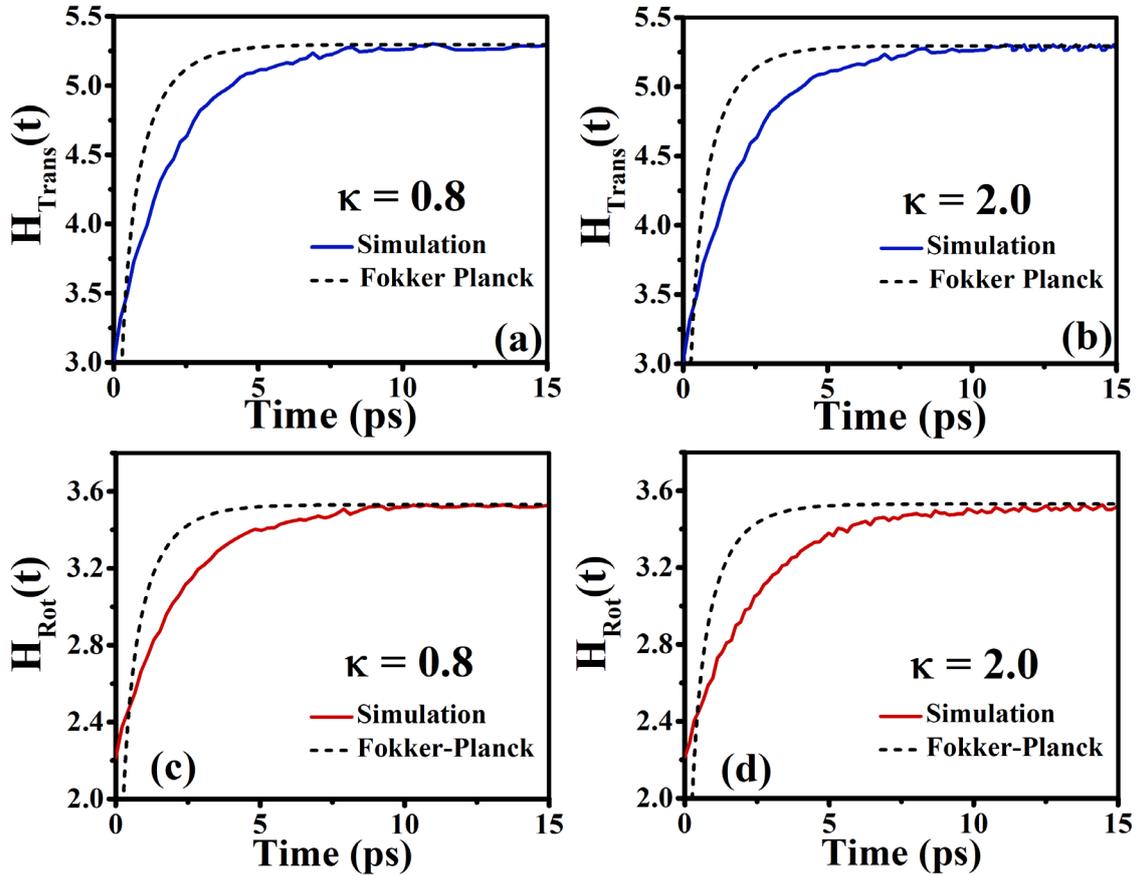

Figure 7: The comparison of translational $H$-function obtained via computer simulation with the analytical solution from the Fokker-Planck equation given by Eq. (17) for (a) oblate ($\kappa = 0.8$) and (b) prolate ($\kappa = 2.0$). Panels (c) and (d) depict the comparison of rotational $H$-function obtained via simulation with the analytical solution from the Fokker-Planck equation given by Eq. (18) for oblate ($\kappa = 0.8$) and prolate ($\kappa = 2.0$), respectively.

Interestingly, the good agreement we observed for the analytical expression for spherical particles is missing here. That is, we do not observe, in the present case, the same level of agreement as obtained for atomistic gases. The reason is not fully understood. We attribute this to the presence of translation-rotation coupling, which is especially strong at intermediate times, making the decomposition of $f(\mathbf{p},\mathbf{L},t)$ into the product of distributions questionable.

Further, the Fokker-Planck theory described above considers separate FP equations for linear and angular momenta. That is, the joint probability distribution is decomposed into two independent probability distributions, one for linear momentum and another for angular momentum. Specifically, we use one FP equation to describe the evolution of the probability distribution function for linear momentum and another separate FP equation for angular momentum. This approach provides simplification and allows us to independently analyze the dynamics of translational and rotational degrees of freedom. However, it is clear that this approach does not explicitly account for the coupling between linear and angular momenta. The possible discrepancies between MD simulations and FP solutions can indeed, and probably do, arise from this lack of coupling in the FP calculations. Real systems often exhibit interactions between translational and rotational degrees of freedom, which are not captured when considering these equations independently.

To address this, a more sophisticated FP approach that includes a single equation encompassing both momenta, thereby capturing their coupling effects (cross terms), could potentially improve the agreement with MD simulations. Developing such an approach is a non-trivial task that requires careful theoretical development and consideration.

## VI. CONCLUDING REMARKS

In this work, we provide a generalization of the famous Boltzmann *H*-function, *H*(*t*), *to include the orientational degree of freedom*. Most of the previous calculations considered only spherical atoms.[40,44] We evaluate the generalized $H(t)$ for an interacting system of molecules in the gas phase from an initial nonequilibrium dynamical state. The calculations of $H(t)$ reveal unique aspects of relaxation of the nonequilibrium distribution function, summarized below.

(i) The orientational relaxation is found to be more sensitive than the translational mode to the aspect ratio of the ellipsoid of revolution, although the time scales are similar.

(ii) The orientational relaxation rate of $H(t)$ increases rapidly as the perfect sphere limit is approached in a manner reminiscent of the slip hydrodynamics boundary condition in rotation.

(iii) The effect of this rapid increase in orientational relaxation rate (as measured by $H(t)$) also shows up in the translational component, which is a new result, due to a coupling between translation and rotational modes, but not fully understood yet. The presence and the magnitude of translation-rotation coupling are captured through a discriminant $\Delta_{T-R}(t)$ defined by Eq. (14). This function shows interesting time dependence.

(iv) We find that a simple Fokker-Planck equation with decoupling between translation and orientation components does not provide a quantitative description of the relaxation function either for the translational or rotational component. This is entirely different from what was observed in an earlier work, where molecules interact with only radial interaction.[44] In that case, the Fokker-Planck equation was found to provide a fairly quantitative agreement of the relaxation except at very short times.

The present study opens the door for further research. A theory is needed to understand the translation-rotation coupling, at least at the level of the Fokker-Planck equation. Although an earlier study by Deutch and Oppenheim[69] addressed this point, no numerical implementation was carried out. This remains a worthwhile project for the future. In fact, earlier studies by Chandra and Bagchi[70] considered the role of translation and rotational density contributions to the rotational and translational friction and discussed the translation-rotation coupling. However, an explicit demonstration of such coupling as demonstrated here was not available earlier.

Next, the present study can also be extended to the study of binary mixtures. In particular, the study of an ellipsoidal particle in a sea of spheres should be studied as this model was considered earlier by Evans and coworkers[53,54] and also Vasanthi *et al.*[71,72] It will be particularly interesting to study the Boltzmann function for a binary mixture of prolates and oblates. It will also be interesting to study $H(t)$ for real molecular gases, like water vapour or carbon dioxide, to ascertain the extent of translation-rotation coupling.

In conclusion, we would like to stress the uniqueness of the present study in understanding the dynamics of molecular liquids, particularly in the gas phase. Evans and coworkers[53,54] earlier derived expressions for the rotational Enskog friction, which can now be used in the Fokker-Planck equation to understand translation-rotation coupling in such liquids. This is a non-trivial problem but deserves further study. The present study is expected to retain some relevance even in denser (gas or liquid) systems where potential energy contribution may dominate the total value of the entropy.

## ACKNOWLEDGEMENTS

We thank Prof. Shinji Saito and Dr. Subhajit Acharya for useful discussions. B.B. thanks the SERB-DST, India, for than India National Science Chair (NSC) Professorship and for providing partial financial support. S.K. thanks the Council of Scientific and Industrial Research (CSIR) for a research fellowship.

# AUTHOR DECLARATIONS

## Conflict of Interest

The authors have no conflicts to disclose.

## DATA AVAILABILITY STATEMENT

The data that supports the findings of this study are available within the article.